\documentclass[final,3p,times,twocolumn]{elsarticle}

\usepackage{amssymb}
\usepackage{amsmath}

\usepackage{natbib}
\usepackage{aas_macros}
\journal{Galaxies}

\begin{document}

\begin{frontmatter}

\title {Time dilation observed in Type Ia supernova light curves and its cosmological consequences}


\author {V\' aclav Vavry\v cuk}

\affiliation{organization={Faculty of Science, Charles University},
            addressline={Albertov 6}, 
            city={Prague},
            postcode={12800}, 
            state={},
            country={Czech Republic}}

\begin{abstract}
The cosmic time dilation observed in Type Ia supernova light curves suggests that the passage of cosmic time varies throughout the evolution of the Universe. This observation implies that the rate of proper time is not constant, as assumed in the standard FLRW metric, but instead is time-dependent. Consequently, the commonly used FLRW metric should be replaced by a more general framework, known as the Conformal Cosmology (CC) metric, to properly account for cosmic time dilation. The CC metric incorporates both spatial expansion and time dilation during cosmic evolution. As a result, it is necessary to distinguish between comoving and proper (physical) time, similar to the distinction made between comoving and proper distances. In addition to successfully explaining cosmic time dilation, the CC metric offers several further advantages: (1) it preserves Lorentz invariance, (2) it maintains the form of Maxwell’s equations as in Minkowski space-time, (3) it eliminates the need for dark matter and dark energy in the Friedmann equations, and (4) it successfully predicts the expansion and morphology of spiral galaxies in agreement with observations. 

\end{abstract}

\end{frontmatter}

\section{Introduction}

The concept of an expanding universe dates back to \citet{Lemaitre1927} and \citet{Hubble1929}, who observed a systematic redshift of nearby galaxies, which was roughly proportional to their distance. This observation, known as the Hubble-Lemaitre law, was initially interpreted as a Doppler effect caused by galaxies moving away from Earth due to the expansion of the Universe. Although this interpretation was later revised, the idea of an expanding universe within the framework of general relativity became a cornerstone of modern cosmology. Today, the cosmological redshift is understood as the effect of the increasing wavelength of photons travelling through expanding space, which is described by the Friedmann-Lemaitre-Robertson-Walker (FLRW) metric  \citep{Weinberg1972, Peacock1999, Carroll2004}
\begin{equation}\label{eq1}
ds^2 = -c^2 dt^2 + a^2\left(t\right) dl^2 \,, \, dl^2 = \frac{dr^2}{1-kr^2}+r^2d\Omega^2 \,, 
\end{equation}
where $ds$ is the spacetime element, $a(t)$ is the scale factor defining cosmic expansion, $c$ is the proper speed of light, $t$ is the cosmic time,  $l$ is the comoving space distance, $k$ is the Gaussian curvature of space, $r$ is the comoving radial distance, and $\Omega$ is the comoving solid angle.

\begin{figure*}
\centering
\includegraphics[angle=0,width=15 cm, clip=true, trim=40 140 40 100]{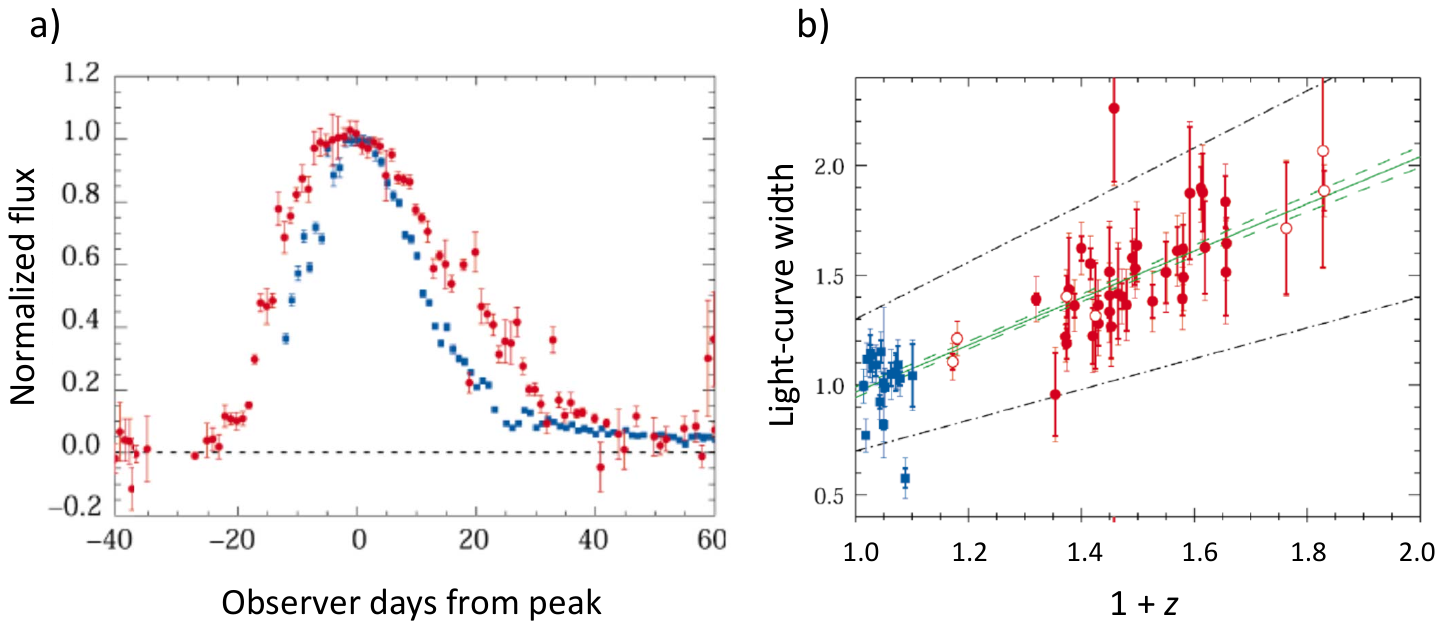}
\caption{
(a) Photometry points for 35 distant (filled red circles) and 18 close (blue squares) SNe averaged over 1 day intervals and over each set of SNe. (b) Observed light curve width versus redshift. Open red circles correspond to another set of 7 fully analyzed SNe. For details, see \citet[their figs 1b and 3a]{Goldhaber2001}. 
}
\label{fig:1}
\end{figure*}

Cosmological redshift is not the only relativistic effect observed in the Universe. Another important effect is cosmic time dilation, detected in Type Ia supernova (SN Ia) light curves. The SN Ia events serve as standard candles in the Universe, making them valuable for probing the cosmic expansion \citep{Riess1998, Perlmutter1999}. Additionally, they act as standard clocks due to their uniform luminosity evolution over time \citep{Leibundgut1996, Leibundgut2001}. Cosmic time dilation, evidenced by the time-stretching of light curves in the observer frame, has been well documented by many authors \citep{Leibundgut1996, Goldhaber1997, Goldhaber2001, Phillips1999, Foley2005}. \citet{Goldhaber2001} analysed SN Ia light curves for 35 high-redshift SNe with $z < 0.8$ discovered by the Supernova Cosmology Project (SCP) and 18 low-redshift SNe with $z < 0.11$. The data were aligned, normalized, and $K$-corrected to establish a common rest-frame $B$-band curve. Comparing the light curves of individual SNe with this reference curve confirmed the presence of cosmic time dilation, with a time-stretch factor of $1 + z$ (see Fig.~\ref{fig:1}).

\begin{figure*}
\centering

\includegraphics[angle=0,width=15 cm, clip=true, trim=40 120 40 100]{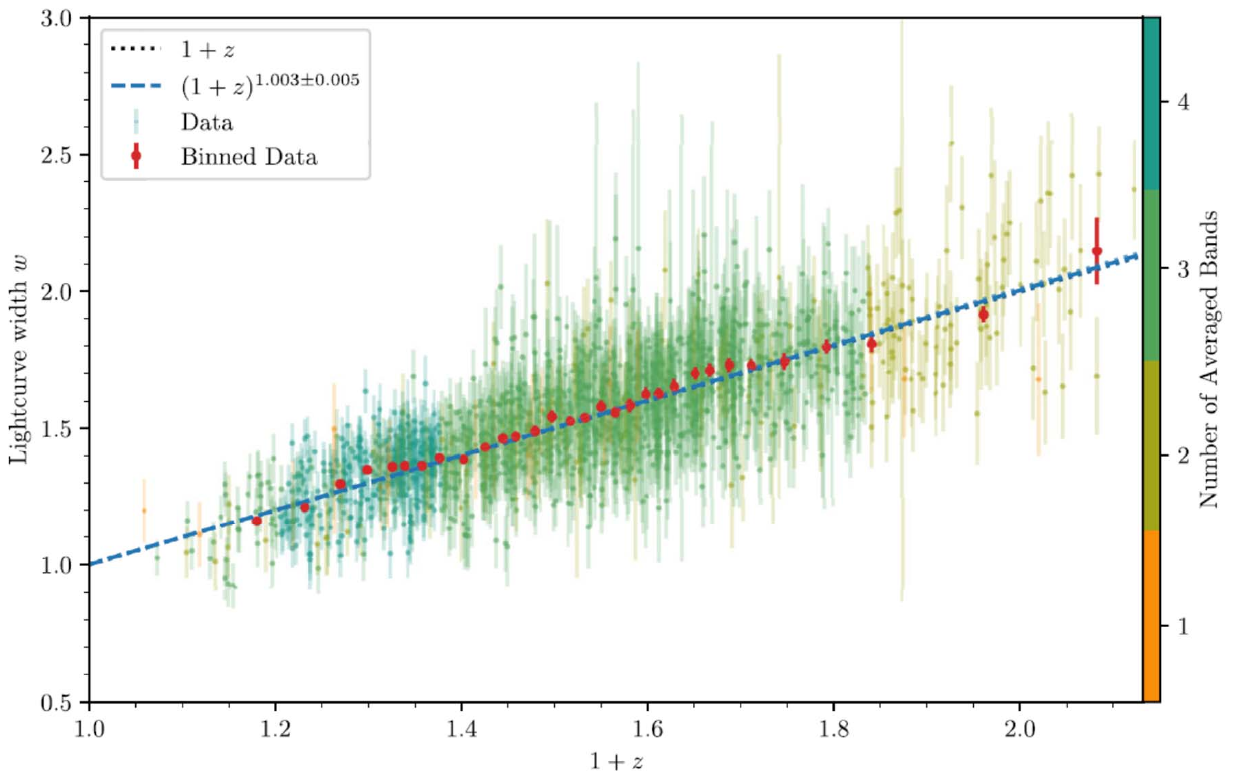}
\caption{
Light curve width as a function of redshift for 1504 SN Ia from the Dark Energy Survey spanning a redshift range of $0.1 \leq z \leq 1.2$. The width value is averaged across four bands. Points are colour-coded according to how many bandpasses were used in averaging. The light curve width increases with redshift as $(1+z)^{1.003 \pm 0.015}$. For details, see \citet[their fig. 8]{White2024}. 
}
\label{fig:2}
\end{figure*}

The time dilation in high-redshift SN Ia was also revealed by \citet{Blondin2008} through determining spectral ages in the supernova rest frame. The authors analyzed 959 spectra of 79 low-redshift $(z < 0.05)$ SN Ia and tested their data for a power-law dependence of the aging rate on redshift. They found that the best-fit exponent for these models was consistent with the expected $1/(1+z)$ factor. Additionally, \citet{White2024} reported a measurement of cosmic time dilation using light curves from 1504 SN Ia with a redshift range of $0.1 \leq z \leq 1.2$ from the Dark Energy Survey. They found that the width of light curves is proportional to $(1+z)^b$ with $b = 1.003 \pm 0.015$ (see Fig.~\ref{fig:2} ). Due to the large sample size, this analysis represents the most precise measurement of cosmic time dilation to date, effectively ruling out non-time-dilating cosmological models with high significance. 
Hence, the systematic stretching of light curves at high redshift is well-established, and corrections for time dilation are now routinely applied to SN Ia data \citep{Leibundgut2001, Goobar_Leibundgut2011}. These observations confirm that the rate of cosmic time varies with the scale factor $a(t)$, see Figs.~\ref{fig:1}b and ~\ref{fig:2}.


Evidence for cosmic time dilation is observed not  only in Type Ia supernova light curves but also in gamma-ray bursts (GRBs), which are among the most energetic explosions in distant galaxies  \citep{Piran2004}. The durations of GRB light curves are highly energy-dependent, making the detection of a time dilation-redshift relation more difficult than in SN Ia. However, this challenge can be overcome by selecting a fixed energy range in the rest frame and measuring the observed durations of GRBs \citep{Sakamoto2011, Ukwatta2012}. Using this approach,  \citet{Zhang2013} analyzed 139 Swift long GRBs with redshift  $z \leq 8.2$ in the energy band from 140 keV to 350 keV and found a correlation between GRB duration and redshift. Similarly, \citet{Littlejohns_Butler2014} studied 232 GRBs detected by the Swift/Burst Alert Telescope (BAT) and found that the observed durations ($T_{90}$, $T_{50}$ and $T_{R45}$) are consistent with cosmic time dilation. Although, GRB observations generally align with cosmic time dilation, this evidence is not conclusive due to uncertainties in the emission mechanisms and limited understanding of GRB physics \citep{Kocevski_Petrosian_2013, Golkhou_Butler_2014, Horvath2022, Singh_Desai_2022}. 

Cosmic time dilation has also been investigated in quasar observations, though the variability of quasar timescales makes the effect more difficult to detect. Some studies have questioned whether time dilation is present in quasar data \citep{Hawkins2001, Hawkins2010}. However, \citet{Dai2012} treated quasars as standard clocks, characterizing light curve segments using slope analysis and finding a correlation between slope and redshift. Nevertheless, the sample of 13 quasars with redshifts of $z \sim 2$, was rather small and the results were statistically inconclusive. A more recent sample from the Dark Energy Survey included 190 quasars with redshifts $z \sim 0.2 - 4.0$ \citep{Stone2022}. \citet{Lewis_Brewer_2023} analysed this dataset and found that quasar variability scales as $(1+z)^n$ with $n =1.28_{-0.29}^{+0.28}$ consistent with time dilation. A follow-up study by \citet{Brewer2025} identified additional scatter in quasar time-scales, refining the result to $n = 1.14 \pm 0.34$, still consistent with cosmic time dilation evolution.

Tests of cosmic expansion described by the FLRW metric are also being conducted through gravitational wave observations, particularly using strongly lensed gravitational and electromagnetic waves. These tests allow for constraints on the curvature parameter of the Universe and provide opportunity to verify the assumptions of homogeneity and isotropy of the Universe underlying the FLRW model \citep{Rasanen2015, Cao2019}. For example, \citet {Qi2019_FLRW_metric} report a spatial curvature measurement of  $\Omega_k = 0.15_{-0.03}^{+0.04}$.

Together, these observations appear to support the validity of the FLRW metric in Eq.~\eqref{eq1} describing an expanding universe within the framework of general relativity (GR). However, a closer inspection of Eq.~\eqref{eq1} and comparison with other metrics describing time dilation effects, such as the Doppler metric in special relativity \citep{Vavrycuk_MPLA_2024} or the Schwarzschild metric in GR  \citep{Weinberg1972, Misner1973, Lambourne2010}, reveals a fundamental shortcoming of the FLRW metric. While this metric accurately describes spatial expansion, it fails to account for cosmic time dilation, including the cosmological redshift. \citet{Melia2019} and \citet{Vavrycuk_Frontiers_2022} showed that observations of cosmic time dilation are inherently tied to the varying time-time term of the metric tensor, $g_{tt}$, known as the lapse function, which evolves with cosmic time. In the standard FLRW metric, $g_{tt} = 1$, meaning the lapse function is constant and independent of cosmic time. As a result, time dilation is not included in this metric  \citep{Melia2019, Vavrycuk_Frontiers_2022, Vavrycuk_JARE_2023, Vavrycuk_Frontiers_Astron_Space_Sci_2023}. Similarly, \citet{Lee2024b, Lee2024c} emphasizes that cosmic time dilation is not predicted by the FLRW metric and must be introduced as an additional assumption.
 
Importantly, setting $g_{tt} = 1$ in the FLRW metric is merely a convention, not a requirement of GR. The theory allows for more general time behaviour, and there is no compelling reason to assume an invariant cosmic time throughout the evolution of the Universe. A well-known counterexample is the Schwarzschild metric, where $g_{tt}$ is not constant but varies, just as the spatial components of the metric tensor do \citep{Vavrycuk_EPJ_Plus_2025}. Thus, the FLRW metric with $g_{tt} = 1$ is simply one admissible solution within GR for describing the evolution of a homogeneous and isotropic universe. While it is the simplest, it is not the most general; other metrics exist in GR that accommodate both spatial expansion and time dilation during cosmic evolution \citep{Endean1994, Endean1997, Mannheim2006, Ibison2007, Kastrup2008, Dabrowski2009, Gron_Johannesen_2011a}.

In this paper, we demonstrate that the cosmic time dilation observed in Type Ia supernova light curves  is successfully explained by the Conformal Cosmology (CC) metric. Unlike the FLRW metric, the lapse function $g_{tt}$ in the CC metric is not 1 but it is proportional to the scale factor $a(t)$. This formulation predicts a time-varying rate of cosmic time, as manifested by observations of cosmic time dilation, and the varying speed of light over cosmic history. We emphasize the importance of distinguishing between comoving and proper (physical) time when studying evolution of the Universe. Finally, we discuss the broader implications of applying the CC metric to fundamental cosmological problems and highlight its potential to resolve key issues in the standard $\Lambda$CDM model, such as the interpretation of the supernova dimming, flat galaxy rotation curves, the dynamics of galaxies and other gravitational systems in an expanding universe.

\section{Cosmological metric with time dilation}

If the rate of cosmic time varies with cosmic evolution as indicated by the SN Ia observations, a more general metric than the FLRW metric must be considered  \citep{Ibison2007, Gron_Johannesen_2011a, Harada2018}
\begin{equation}\label{eq2}
ds^2= -g_{tt} c^2 dt^2 + g_{ll} dl^2 = -A^2(t) c^2 dt^2 + B^2(t) dl^2 \,, 
\end{equation}
where $ds$ is the spacetime  element, $g_{tt}$ and $g_{ll}$ are the time and space components of the metric tensor, $c$ is the comoving (contravariant) speed of light, $t$ is the comoving (contravariant) time, $l$ is the comoving (contravariant) space distance, and $A(t)$ and $B(t)$ are the lapse and expansion functions, which are arbitrary functions that describe time dilation and space expansion, respectively.

The simplest way to define a metric that describes both cosmic expansion and time dilation is to assume that $A(t)=B(t)=a(t)$: 
\begin{equation}\label{eq3}
ds^2 = -a^2\left(t\right) c^2 dt^2 + a^2\left(t\right) dl^2 = a^2\left(t\right) \left(-c^2 dt^2 +  dl^2 \right) \,, 
\end{equation}
where $a(t)$ is the scale factor defining cosmic expansion. 

It is important to emphasize that the time $t$ in  Eq.~\eqref{eq3} refers to comoving cosmic time, but not `conformal time' $\tau$. Conformal time $\tau$ is often used in cosmology as a parameter which has no direct physical meaning. It is obtained by artificial rescaling proper time $T = t$ in the FLRW metric in Eq.~\eqref{eq1}. Hence, the conformal time $\tau$ is a result of a coordinate transformation in the FLRW metric with no physical consequences, keeping the lapse function $g_{tt}$ equal to 1 \citep{Islam2001}. In the metric described by Eq.~\eqref{eq3}, the time $t$ is the true comoving time, and the lapse function $g_{tt} = a(t)$ varies with time.

The metric in Eq.~\eqref{eq3} is known as the conformal cosmology (CC) metric  \citep{Endean1994,Endean1997,Kastrup2008,Dabrowski2009,Gron_Johannesen_2011a} and has exceptional properties. Firstly, it evolves according to the conformal transformation, a concept well-studied in general relativity \citep{Mannheim2006, Capozziello_deLaurentis2011, Penrose2011a}. Secondly, this metric is Lorentz invariant and preserves the form of Maxwell's equations in Minkowski spacetime \citep{Infeld_Schild_1945, Infeld_Schild_1946, Ibison2007}. 

\subsection{Comoving and proper time}
Based on physical arguments, we assert that the FLRW metric is inconsistent with SN Ia observations of cosmic time dilation, as it does not predict a varying rate of cosmic time. To support this claim, we mathematically derive the relation between comoving and proper (physical) times in the FLRW and CC metrics within the framework of curvilinear coordinate formalism \citep{Arfken_Weber_1995,Hartle2003,Cook2004}. 

The comoving coordinates $t$ and $l$ in Eq.~\eqref{eq3} are contravariant quantities, meaning they are not coordinate-invariant and do not represent true physical cosmic time or distance. The reason is straightforward: in curvilinear coordinate systems, the base vectors are generally non-unit. True physical quantities are the proper time $T$ (measured by atomic clocks) and the proper distance $L$ (measured by a rigid rod). These quantities are coordinate-invariant, meaning they are independent of the choice of the coordinate system. They are calculated either by taking the product of covariant and contravariant time and distance coordinates or by employing an orthonormal coordinate basis, as explained by \citet{Hartle2003} or \citet{Cook2004}. Therefore, the proper time $T$ and proper distance $L$ are given by (see Appendix A, Eq.~\eqref{eqA8})
\begin{equation}\label{eq4}
dT = \sqrt{g_{tt}}\, dt\,,\,\, dL = \sqrt{g_{ll}}\, dl \,. 
\end{equation}

\begin{figure*}
\centering

\includegraphics[angle=0,width=15 cm, clip=true, trim=40 120 40 100]{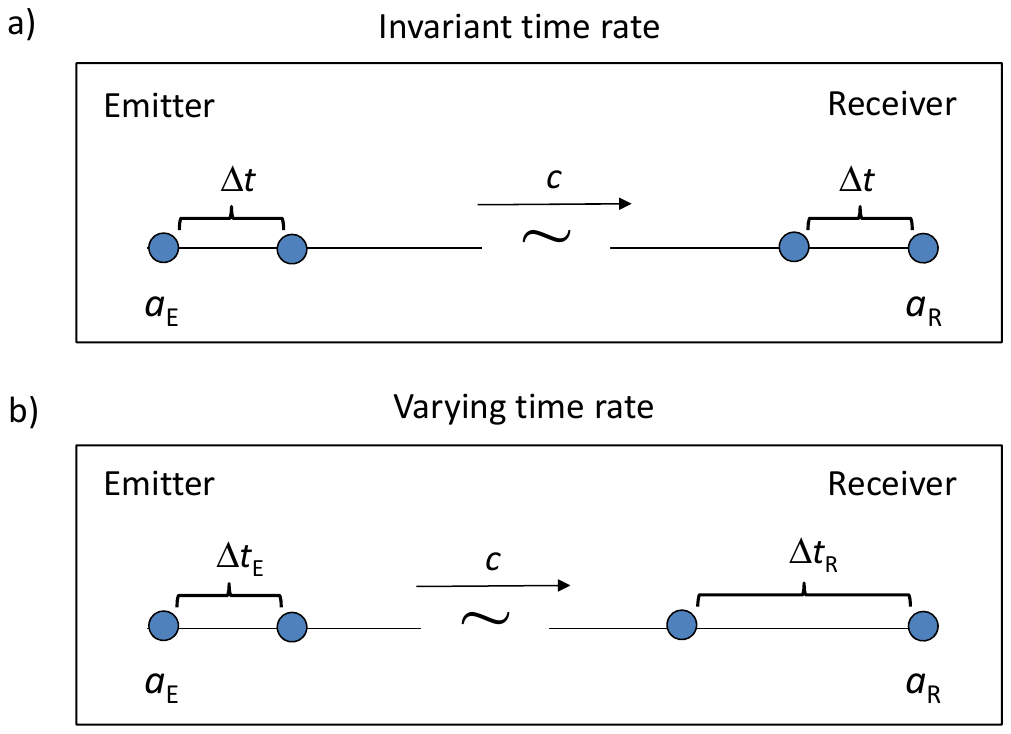}
\caption{
Schematic representation of photons (blue filled circles) propagating from the emitter to the receiver. (a) Metric with a fixed time rate (see Eq.~\eqref{eq1}). (b) Metric with a varying time rate (see Eq.~\eqref{eq3}). The quantity $a$ represents the scale factor, $\Delta t$ denotes the time rate, and $c$ is the speed of light. Subscripts 'E' and 'R' indicate values at the emitter and receiver, respectively. 
}
\label{fig:3}
\end{figure*}

It follows from Eq.~\eqref{eq4} that Eqs.~\eqref{eq1} and ~\eqref{eq3} describe two physically different models of the Universe. Although both the FLRW and CC metrics describe space expansion in a similar manner: 
\begin{equation}\label{eq5}
dL = a(t) dl \,, 
\end{equation}
they treat time differently. In the FLRW metric
\begin{equation}\label{eq6}
dT = dt \,, 
\end{equation}
implying that proper time is independent of expansion (Fig.~\ref{fig:3}a). In contrast, in the CC metric:
\begin{equation}\label{eq7}
dT = a(t) dt \,, 
\end{equation}
implying that proper time depends on expansion (Fig.~\ref{fig:3}b). Therefore, in the FLRW metric, comoving and proper times are equivalent, whereas in the CC metric, they are not. 

The difference in the evolution of cosmic time between the two metrics provides further evidence that the FLRW and CC models describe physically distinct universes. Importantly, the CC metric accounts for both space expansion and time dilation during the evolution of the Universe, which is essential for satisfactorily explaining observations of  cosmic time dilation in the SN Ia light curves (see Figs.~\ref{fig:1} and ~\ref{fig:2}). 


\subsection{Comoving and proper speed of light}
The fundamental difference between the FLRW and CC metrics is clearly demonstrated through the properties of the comoving (contravariant) and physical (proper) speeds of light in both models. 

The propagation of light is described by the null geodesics equation, $ds^2 = 0$. For the FLRW metric, this equation takes the form: 

\begin{equation}\label{eq8}
c^2 dt^2 = a^2(t) dl^2 \,. 
\end{equation}
Consequently, the comoving light speed  $\hat{c}$ is given by:
\begin{equation}\label{eq9}
\hat{c} = \frac{dl}{dt} = \frac{c}{a(t)} \,,
\end{equation}
where $t$ is the affine parameter describing the null worldline.

Since the basis vectors defining the FLRW metric are not unit vectors, the comoving (contravariant) and true physical speeds of light differ. The comoving speed of light depends on the choice of the coordinate system because it is evaluated in a non-orthonormal vector basis. To determine the physical (proper) speed of light $C$, which is coordinate-invariant and measurable in physical experiments (e.g., using atomic clocks and rigid rods), we must use the orthonormal coordinate basis (see Appendix A, Eq.~\eqref{eqA8})
\begin{equation}\label{eq10}
C = \sqrt{g_{ii}} \, \hat{c} = a(t) \hat{c} = c \,.
\end{equation}
Thus, we find that the physical speed of photons $C$ remains constant, $C = c$, in the FLRW metric independent of cosmic time. 

\begin{figure*}
\centering
\includegraphics[angle=0,width=12 cm, clip=true, trim=180 110 120 100]{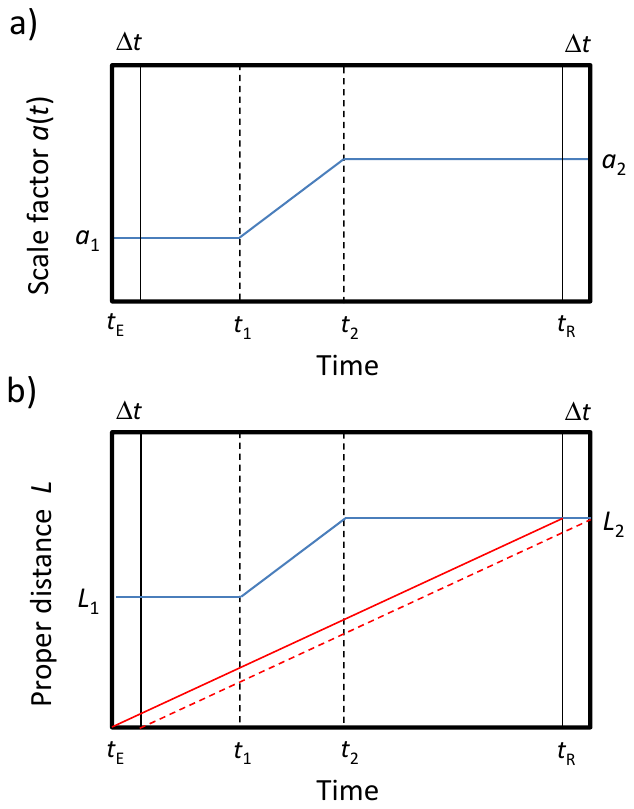}
\caption{
Photons travelling in an expanding universe described by the FLRW metric. (a) Example of the expansion function $a(t)$, where expansion occurs between times $t_1$ and $t_2$. (b) Proper distance $L$ between the emitter and receiver as a function of time (blue line) and proper travel distance $L$ of two successive photons moving in the expanding space (red lines). The solid red line -- the first photon, the dashed red line -- the second photon. Since both photons travel at the same velocity $c$, the time delay $\Delta t$ between them remains constant along the entire common ray path.
}
\label{fig:4}
\end{figure*}

The constant speed of light has significant implications for photons travelling across the expanding universe: their mutual distance and time delay remain unchanged over cosmic time (see Fig.~\ref{fig:4}). This statement contradicts the common belief that the distance between the two successive photons must increase over time in an expanding universe described by the FLRW metric. This idea was first introduced by \citet{Lemaitre1927} and later reiterated in many textbooks \citep{Weinberg1972, Misner1973, Islam2001, Hartle2003, Lambourne2010}. The fundamental issue with Lemaitre's derivation \citep{Lemaitre1927} lies in the incorrect definition of wavelength as the distance between two different spacetime events. Importantly, distance must be measured within a single coordinate system, rather than as the distance between points in two different coordinate systems associated with two photons observed at different times. When considering two photons travelling along the same ray path with an initial proper distance $d$ between them at the same coordinate time, the effect of increasing distance between photons due to space expansion vanishes. Since both photons traverse the same path at the same speed, their mutual distance remains constant over time. Consequently, while the distance between massive objects at rest increases in the expanding universe, the separation between photons travelling along the same ray path does not change \citep[his Appendix B and C]{Vavrycuk_Frontiers_2022}. 

The conclusion is clearly illustrated in Fig.~\ref{fig:4}, where the time-dependent expansion is specifically chosen such that the Universe is not expanding at the moments when the photons are emitted and received. Thus, it is evident that the length of the ray paths of both photons, traversing the distance between the emitter and receiver, must be exactly the same. Given the constant speed of light, their travel time must also be identical.


It is important to note that the physical speed of light $C$ cannot be calculated as the ratio of the proper distance element $dL$ to the proper cosmic time element $dT$, i.e., $C \neq dL/dT$ , as is commonly assumed. Instead, it must be determined as the ratio of the proper distance element $dL$ to the physical travel time element of light $d\tau$, such that $C = dL/d\tau$. If we mistakenly calculate the speed of light as $dL/dT$ and use Eqs.~\eqref{eq5} and ~\eqref{eq6} for $dL$ and $dT$, we obtain an increasing proper speed of light over cosmic time in the FLRW metric:
\begin{equation}\label{eq11}
\frac{dL}{dT} = a(t) \frac{dl}{dt} = a(t) c \,,
\end{equation}
which is incorrect. The error in this intuitive approach arises from confusing the  \textit{physical travel time} element $d\tau$ with the \textit{physical cosmic time} element $dT$ in Eq.~\eqref{eq11}. Clearly, if galaxies are receding and the speed of light is constant, the travel time $\tau$ of light between galaxies increases. However, the rate of proper cosmic time remains constant. Therefore, these two time elements should not be interchanged.

For the CC metric, the null geodesic equation, $ds^2 = 0$, yields
\begin{equation}\label{eq12}
a^2(t) \left(-c^2dt^2 + dl^2\right) = 0 \,. 
\end{equation}
Hence, the comoving (contravariant) speed of light $\hat{c}$ is
\begin{equation}\label{eq13}
\hat{c} = \frac{dl}{dt} = c \,.
\end{equation}

Similarly to the FLRW metric, the physical speed of light $C$ in the CC metric must be calculated using the orthonormal coordinate basis (see Appendix A, Eq.~\eqref{eqA8})
\begin{equation}\label{eq14}
C = \sqrt{g_{ll}} \, \hat{c} = a(t) \hat{c} = a(t) c \,.
\end{equation}
Consequently, the physical speed of light $C$ increases during the cosmic expansion in the CC metric. While this may seem counter-intuitive, it is physically reasonable. In the CC metric, as the distances between galaxies increase, time dilation also increases. These effects compensate for each other, ensuring that the total travel of light between galaxies remains constant. As a result, the speed of light must increase over cosmic time to maintain the same travel duration.

\section{Physical differences between the FLRW and CC metrics}

Eqs.~\eqref{eq10} and ~\eqref{eq14} demonstrate that the FLRW and CC metrics are physically distinct: 
In the FLRW metric, the physical speed of light remains constant over cosmic history, $C = c$, while in the CC metric, the physical speed of light varies with cosmic time, $C = a(t)c$.
	Similarly, the CC metric and the Minkovski metric are physically distinct. Although the CC metric is conformal to the Minkowski metric, their physical properties differ:
In the Minkowski metric, the time rate is constant, space is static, and the physical speed of light remains constant, 
while i	n the CC metric, the time rate varies, space is expanding, and the physical speed of light also varies.

Emphasize that a varying speed of light (VSL), as predicted by the CC metric, is not excluded within GR. Since the gravitational field of the Universe evolves with cosmic expansion, it is natural to expect that the speed of light also varies with cosmic time. As stated by \citet{Einstein1920}: 
'\textit{The law of the constancy of the velocity of light in vacuo, which constitutes one of the fundamental assumptions in the special theory of relativity and to which we have already frequently referred, cannot claim any unlimited validity}' ... ' \textit{its results hold only so long as we are able to disregards the influences of gravitational fields on the phenomena (e.g., of light)}'.

To avoid confusion regarding the differences between the CC, FLRW and Minkowski metrics, it is important to emphasize the following key points:

\begin{itemize}
\item
\textit{Physical meaning of metrics.} All vector and tensor quantities in curvilinear coordinate systems are coordinate-dependent. They cannot be directly interpreted in physical terms because their basis vectors are not orthonormal. In cosmology, the primary purpose of evaluating the metric tensor is to express spacetime in coordinates that can be simply translated into physically meaningful quantities. These quantities must always be coordinate-invariant and should be expressed using the orthonormal tetrad of basis vectors (see Appendix A). 
\item
\textit{Misleading equivalence between rescaled metrics.} A common belief is that the FLRW and CC metrics are physically equivalent because one can be transformed into the other through rescaling or time synchronization (\citep[his eq. 11.83]{Weinberg1972}, \citep[ their eq. 27.14]{Misner1973}). This is misleading. Although Einstein's field equations are coordinate-invariant, arbitrarily rescaling components of the metric tensor may have physical consequences. If such a transformation alters physical units (i.e., changes coordinate-invariant quantities), the resulting metrics describe physically different cosmological models.  
\item
\textit{Expanding vs. static universe.} The metric of an expanding universe can be formally transformed into the metric of a static universe by introducing conformal distance. Although this rescaling is mathematically valid, this transformation does not eliminate the physical distinction between an expanding and a static universe. Similarly, the transformation of a model with a varying time rate into one with a fixed time rate can be performed through introducing conformal time. However, this transformation does not remove the underlying physical differences between the two models.
\item
\textit{Appropriate cosmological model.} Since astronomical observations support both the expansion of space and cosmic time dilation, an appropriate cosmological model should be described by a metric tensor in which both the lapse function $g_{tt}$ and the spatial components $g_{ll}$ vary with time. This model is referred to as the 'Cosmological Coordinate System (CCS)', in which all major astronomical bodies remain at rest \citep{Endean1994, Infeld_Schild_1945, Infeld_Schild_1946}. The metric must also reflect that the clock rates associated with these fundamental bodies vary over cosmic time. 
\end{itemize}

\section{Physical origin of cosmic time dilation}

As discussed in the previous sections, the expansion of the Universe alone does not cause cosmic time dilation. Expansion influences the distances between galaxies at rest with respect to the cosmological reference frame. However, the distance between two successive photons traveling along the same ray path remains unchanged during the expansion. Therefore, the origin of cosmic time dilation must lie in a different physical mechanism.

In general relativity, time distortion is attributed to the presence of a gravitational field. The most well-known experimentally confirmed case is the gravitational redshift of photons due to Earth's gravity, as measured by \citet{Pound_Rebka_1960} and \citet{Pound_Snider_1964}. These experiments demonstrated that the change in photon angular frequency $\Delta \omega$ between two observation points (emitter and receiver) depends on the difference in the Newtonian gravitational potential $\Delta \Phi$ between them, expressed as \citep{Misner1973}
\begin{equation}\label{eq15}
\frac{\Delta \omega}{\omega} = \frac{g z}{c^2} = \frac{\Delta \Phi}{c^2} \,,
\end{equation}
where $g$ is the gravitational acceleration, and $z$ is the height difference between the two points. 

Since a change in photon frequency indicates a change in the rate of time, this variation is characterised by the lapse function  $g_{tt}$ of the metric tensor, such that 
\begin{equation}\label{eq16}
\frac{\omega(E)}{\omega(R)} = \sqrt{\frac{g_{tt}(R)}{g_{tt}(E)}}  \,.
\end{equation}

In weak gravitational fields, where $\Delta \Phi/c^2 \ll 1$, the lapse function $g_{tt}$ can be directly approximated using  \citep[his eq. 4.6]{Vavrycuk_EPJ_Plus_2025} 
\begin{equation}\label{eq17}
g_{tt} = 1 + 2\frac{\Delta \Phi}{c^2}  \,.
\end{equation}
Hence, the lapse function $g_{tt}$ varies as a function of the gravitational potential difference $\Delta \Phi$.

Clearly, a similar physical mechanism should also govern the lapse function $g_{tt}$ in the cosmological metric tensor, since the gravitational field cannot be avoided in cosmology, being generated by the mass-energy content of the Universe itself. As the Universe expands, its total gravitational potential must change over time. 

Given that the Newtonian gravitational potential produced by a massive body depends on the proper radial distance $R$ from this body as $1/R$, and that all distances increase with the scale factor $a(t)$, the total gravitational potential of the Universe $\Phi (t)$ at any point must decrease over time as
\begin{equation}\label{eq18}
\Phi(t) = \frac{1}{a(t)} \Phi_0  \,,
\end{equation}
where $\Phi_0$ is the present-day total gravitational potential of the Universe. 

By analogy with Eq.~\eqref{eq15}, the time-dependent gravitational potential $\Phi(t)$ in Eq.~\eqref{eq18}, should lead to a varying angular frequency $\omega(t)$ of photons propagating in the expanding universe. Accordingly, Eq.~\eqref{eq16} implies that the cosmological lapse function $g_{tt}(t)$ must also vary over time. Thus, both $\omega(t)$ and $g_{tt}(t)$ are functions of the scale factor $a(t)$.

We conclude that the gravitational redshift and the cosmological redshift share the same physical origin: both result from the presence of a gravitational field. In earlier cosmic epochs, the Universe was denser, and the gravitational field was stronger. Therefore, going backward in time from low to high redshift is analogous to approaching a black hole, in the vicinity of which time dilation becomes more significant. Just as time runs differently near and far from a black hole, a similar effect should be expected when comparing clocks in the high redshift Universe and the present epoch. The only difference is that gravitational redshift depends on spatial coordinate, while cosmological redshift depends on cosmic time.

\section{Discussion}
 
The FLRW metric appears to be supported by a wide range of astrophysical and cosmological observations. However, this conventional view focuses primarily on confirming the spatial expansion of the Universe as described by the scale factor $a(t)$. In contrast, the other key effect predicted by general relativity -- time distortion -- is neglected in the FLRW metric, as it assumes a constant lapse function throughout cosmic evolution. Importantly, a rigorous mathematical analysis of Einstein's field equations suggests that this assumption is not valid. Spatial expansion alone cannot cause a change in photon frequency or alter the temporal properties of physical processes in the Universe. This leads to the conclusion that observations of cosmological redshift and cosmic time dilation are inconsistent with predictions of the FLRW metric \citep{Vavrycuk_Frontiers_2022}.

\begin{figure*}
\centering

\includegraphics[angle=0,width=15cm,trim=60 200 50 100]{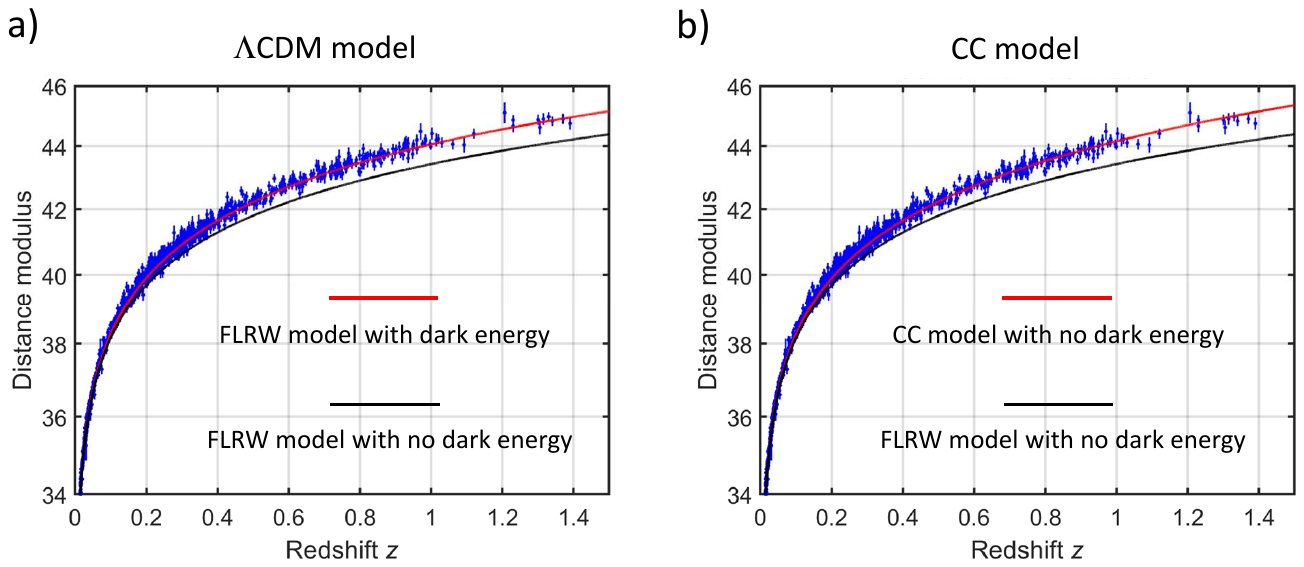}
\caption{
The Hubble diagram with Type Ia supernovae observations. Blue dots show measurements of the SNe Pantheon compilation \citep{Scolnic2018, Jones2018}. The red line in (a) shows the $\Lambda$CDM model based on the FLRW metric with dark energy. The red line in (b) shows the model based on the CC metric without dark energy. The black line in (a-b) shows the standard cosmological model based on the FLRW metric without dark energy. For details, see \citep{Vavrycuk_Frontiers_2022}.
}
\label{fig:5}
\end{figure*}

Because the standard $\Lambda$CDM model is built on the FLRW metric, its well-known tensions \citep{Kroupa2012, Kroupa2015, Buchert2016, Bullock_Boylan-Kolchin2017, Ezquiaga_Zumalacarregui2017, Del-Popolo_Le-Delliou2017} may stem from the FLRW metric's inability to account for all relativistic aspects of an expanding universe. These tensions include: (1) the need to introduce dark energy to explain the dimming of Type Ia supernovae, (2) the need for dark matter to account for observations of flat galaxy rotation curves, (3) inadequacies in describing the expansion and morphology of galaxies, and (4) difficulties in modelling the expansion of local gravitational systems.

By contrast, the Conformal Cosmology metric appears to resolve many of these issues. Most notably, it eliminates the need for dark energy in the Friedmann equations. The speculative and unphysical concept of dark energy was originally introduced into the Friedmann equations to reconcile predictions of the $\Lambda$CDM model (based on the FLRW metric) with the observed dimming of Type Ia supernova, as reported by  \citet{Riess1998} and \citet{Perlmutter1999}. Some alternative approaches have attempted to avoid invoking dark energy by modifying the relation between cosmological redshift and scale factor \citep{Tian2017, Benedetto2024}. 
However, this issue is naturally resolved within the CC framework. Since the Friedmann equations take a different form in the CC metric \citep{Vavrycuk_Frontiers_2022}, they align naturally with Type Ia supernovae luminosity data, without the need for dark energy or modifying the redshift-scale factor relation (see Fig.~\ref{fig:5}). Consequently, the supernova dimming discovered by  \citet{Riess1998} and \citet{Perlmutter1999} does not imply the existence of dark energy or complex redshift-scale factor relationship. Instead, it highlights a fundamental flaw in the FLRW metric \citep{Vavrycuk_Frontiers_2022}.

\begin{figure*}
\centering

\includegraphics[angle=0,width=15 cm, clip=true, trim=60 155 50 160]{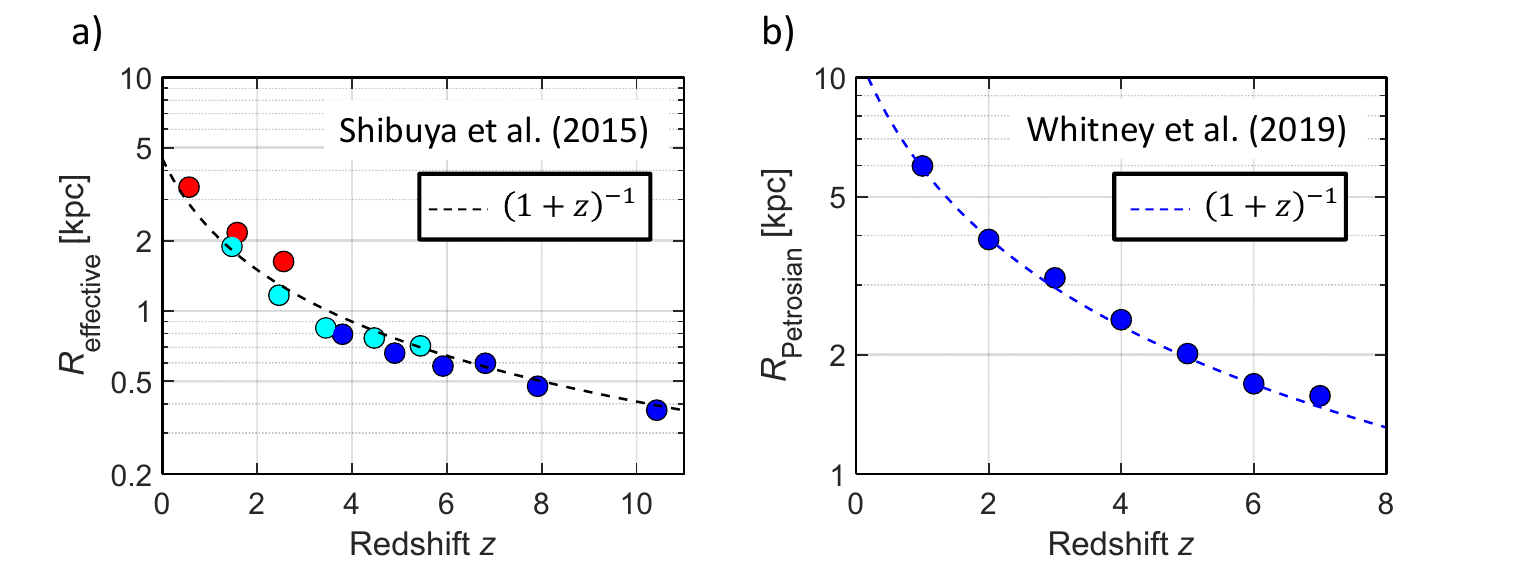}
\caption{
Galaxy size evolution with redshift. (a) Median of the effective galaxy radius $R$ as a function of redshift. The red and cyan filled circles indicate radius $R$ for star-forming galaxies measured in the optical ($4500 - 8000 \,$\AA) and UV ($1500 - 3000 \,$\AA) wavelength ranges, respectively. The blue filled circles indicate radius $R$ for the Lyman break galaxies measured in the UV  wavelength range. For details, see \citet[their fig. 8]{Shibuya2015}. (b) Median Petrosian radius of galaxies as a function of redshift for the mass-limited sample in the range of $10^9 \, \mathrm{M_\odot} \le M_* \le 10^{10.5} \, \mathrm{M_\odot}$.  For details, see \citet [their fig. 8]{Whitney2019}. The dashed lines in (a,b) show the size evolution predicted by the CC metric. Note that the FLRW metric predicts no size evolution.
}
\label{fig:6}
\end{figure*}

\begin{figure*}
\centering

\includegraphics[angle=0,width=15 cm, clip=true, trim=80 200 50 120]{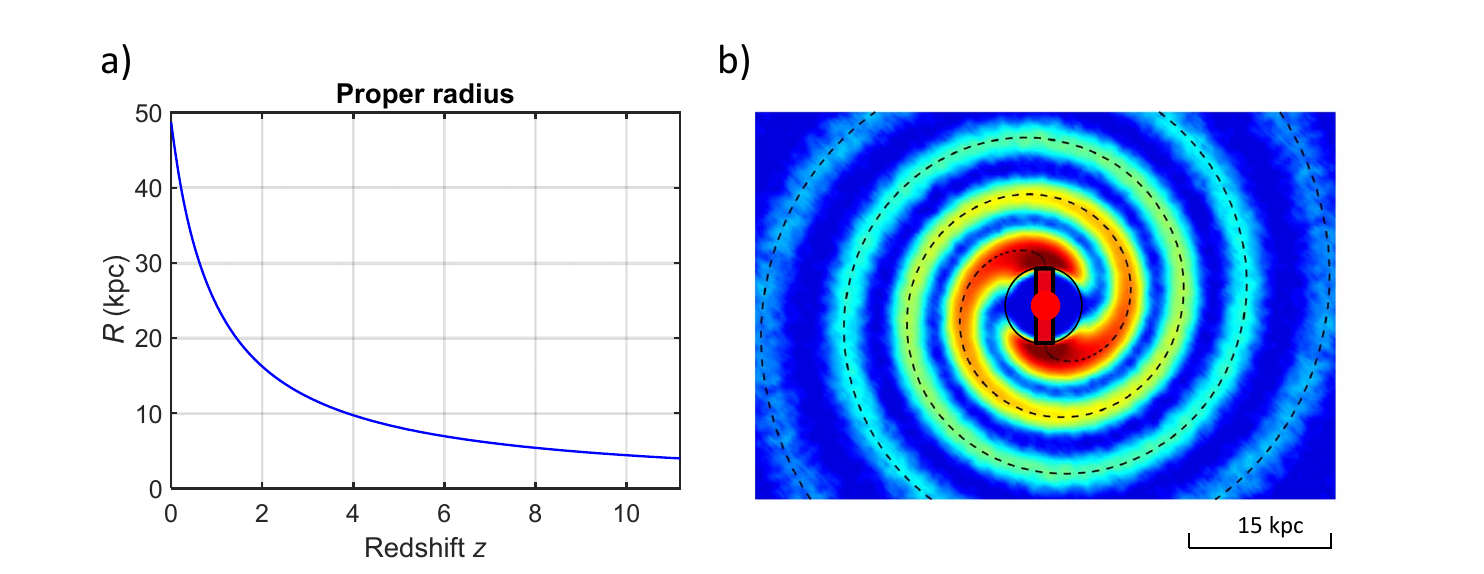}
\caption{
The proper radius of a galaxy as a function of redshift (a) and the spiral arms formed during the galaxy evolution (b) in the CC model. The density of the material in the spiral arms in (b) is colour coded. The red bulge and bar inside the black circle in (b) illustrate schematically the orientation of the bar. For details, see \citet{Vavrycuk_Frontiers_Astron_Space_Sci_2023}.
}
\label{fig:7}
\end{figure*}

Additionally, the CC metric predicts a fundamentally different behaviour of gravitational orbits in an expanding universe compared to predictions based on the FLRW metric. Contrary to the common assumption that local systems resist cosmic expansion, the CC metric predicts that such systems do expand according to the Hubble flow (see Fig.~\ref{fig:6}). According to \citet{Vavrycuk_Frontiers_Astron_Space_Sci_2023}, the proper velocity of massive particles remains constant, regardless of cosmic expansion. This principle also applies to the rotational velocities of stars in galaxies. As galaxies expand, stars gradually drift outward from the galaxy centre to the periphery, while maintaining their rotational speeds. This dynamic naturally produces flat rotation curves, as modelled by \citet{Vavrycuk_Frontiers_Astron_Space_Sci_2023}, without requiring dark matter halos around galaxies. This result challenges the standard interpretation of galaxy dynamics and offers a compelling alternative explanation based on the CC metric. As a consequence, the sizes of spiral galaxies increase over time, a trend that is consistent with observations \citep{vanDokkum2010, Williams2010, Shibuya2015, Whitney2019}, see Figs.~\ref{fig:6} and ~\ref{fig:7}a.

The CC metric also offers a natural explanation for the characteristic spiral structure of galaxies \citep{Vavrycuk_Frontiers_Astron_Space_Sci_2023}. The formation of spiral arms is conventionally attributed to instabilities in the stellar disk due to self-gravity, being modelled by the density-wave theory \citep{Shu1970, Roberts1975, Sellwood_Carlberg_1984, Elmegreen1999, Sellwood2011}. This theory is a primary tool for studying the gravitational stability of disk galaxies. However, the density-wave theory suffers from several open questions and limitations. It is based on the Newtonian gravity, neglects the space expansion, and does not account for the growth of spiral galaxies over cosmic time. When space expansion and time dilation are incorporated into galactic dynamics via the CC metric, the galaxy naturally increases in size, and the spiral pattern emerges as a result \citep{Vavrycuk_Frontiers_Astron_Space_Sci_2023}, (see Fig.~\ref{fig:7}b). Thanks to these general relativistic effects, the spiral structure is not destroyed by the well-known winding problem, a long-standing challenge in galaxy formation models \citep{Ferreras2019}.

Obviously, despite the many advantages of the CC metric over the FLRW metric, further observational evidence is essential to strengthen our results and ultimately to consider replacing the $\Lambda$CDM model with the CC model in cosmological applications. Future research should focus primarily on detecting the hypothetical variation in the speed of light, as predicted by the CC metric and other alternative approaches \citep{Magueijo2003, Moffat2016, Lee2024b, Lee2024c, Santos2024}. This contrasts with the FLRW metric, which assumes a constant speed of light independent of cosmic evolution. In addition, more compelling evidence of cosmic time dilation in GRBs and high-redshift quasars would further reinforce the legitimacy of the CC model. Observations of the expansion of galaxies and other local gravitational systems would also provide strong support for the CC metric framework. This includes a careful reanalysis of the Solar system's dynamics, where the standard model faces persistent challenges such as: the Faint young Sun paradox, the anomalies in the Moon's and Titan's orbital evolution, and the formation and structure of the Kuiper Belt \citep{Krizek2012, Krizek_Somer_2015, Dumin2015}.

\section{Conclusions}
We have shown that the assumption of invariant time, as embedded in the FLRW metric, is inconsistent with observed cosmic time dilation in Type Ia supernova light curves. While this conclusion may seem surprising, it is physically well justified. If clocks at rest ticked at the same rate regardless of cosmic evolution, and the physical speed of light remined constant, then SN Ia light curves would not exhibit temporal stretching at high redshifts. Instead, their durations would appear unchanged, contradicting observations. 

Cosmic time dilation is, in fact, a direct manifestation of a time-varying lapse function $g_{tt}$, similar to the time dilation observed near a Schwarzschild black hole, where it appears as gravitational redshift \citep{Vavrycuk_EPJ_Plus_2025}. Therefore, accurate cosmological models must incorporate the CC metric, in which $g_{tt}$ varies with time. Both cosmic time dilation and gravitational redshift arise from the same physical mechanism: a changing rate of cosmic time and a varying physical speed of light due to spacetime distortion caused by a gravitational field. 

Just as clocks run at different rates depending on their distance from a black hole, clocks also tick differently in the early Universe (high-redshift) compared to the late Universe (low redshift). At high redshifts, the Universe was more compact, resulting in a the stronger gravitational field and greater spacetime distortion compared to low-redshift Universe. As a  consequence, the physical speed of light was lower in the early Universe than it is today. 

The CC metric also offers several additional advantages: (1) it is Lorentz invariant, (2) it preserves the standard form of Maxwell's equations, consistent with their form in Minkowski spacetime, (3) it eliminates the need for speculative concepts of dark matter and dark energy in cosmological models, and (4) it is consistent with observations of galaxy expansion and the behaviour of local gravitational systems.

\section{Acknowledgements}
I sincerely thank three anonymous reviewers for their detailed and thoughtful comments, which have helped to significantly improve the manuscript.




\renewcommand{\theequation}{A-\arabic{equation}}
\setcounter{equation}{0}  

\section*{Appendix A: Riemannian manifold and curvilinear coordinate systems}

Let us assume that $(x^0,x^1,x^2,x^3 )$ is a specific choice of the coordinate system, which covers the Riemannian manifold. These coordinates will be unique and differentiable functions of the Cartesian coordinates $(y^0,y^1,y^2,y^3 )$, covering the Euclidean space $\mathbf{R}^4$. Geometry of the Riemannian manifold is then defined by the covariant and contravariant base vectors $\mathbf{g}_\mu$ and $\mathbf{g}^\mu$ \citet[his eq. 20.43]{Hartle2003}  
\begin{equation}\label{eqA1}
\mathbf{g}_\mu =  \frac{\partial{y^\beta}}{\partial{x^\mu}} \, \mathbf{i}_\beta \,,\,\, \mathbf{g}^\mu =  \frac{\partial{x^\beta}}{\partial{y^\mu}} \, \mathbf{i}^\beta   \,,
\end{equation}
and by covariant and contravariant metric tensors $g_{\mu \nu}$ and $g^{\mu \nu}$  \citep[his eq. 20.44]{Hartle2003}:
\begin{equation}\label{eqA2}
g_{\mu \nu} = \mathbf{g}_\mu \cdot \mathbf{g}_\nu =   
\frac{\partial{y^\alpha}}{\partial{x^\mu}} \frac{\partial{y^\beta}}{\partial{x^\nu}} \, \eta_{\alpha \beta} \,, \,\, g^{\mu \nu} = \mathbf{g}^\mu \cdot \mathbf{g}^\nu =   
\frac{\partial{x^\alpha}}{\partial{y^\mu}} \frac{\partial{x^\beta}}{\partial{y^\nu}} \, \eta^{\alpha \beta} \,,
\end{equation}
where $\mathbf{i}_\beta = \mathbf{i}^\beta$ are the unit Cartesian base vectors in the Minkowski space, and $\eta_{\alpha \beta} = \eta^{\alpha \beta} = \mathrm{diag}(-1,1,1,1)$ is the Minkowski metric. In contrast to the base vectors $\mathbf{i}_\beta$, which are unit in length, the base vectors $\mathbf{g}_\mu$ and $\mathbf{g}^\mu$ are generally non-unit. Vector $\mathbf{v}$ and tensor $\mathbf{T}$ in curvilinear coordinates $x^\alpha$ are expressed as
\begin{equation}\label{eqA3}
\mathbf{v} = v_\alpha \mathbf{g}^\alpha = v^\alpha \mathbf{g}_\alpha  \,,
\end{equation}
and
\begin{equation}\label{eqA4}
\mathbf{T} = T_{\alpha \beta} \, \mathbf{g}^\alpha \mathbf{g}^\beta = T^{\alpha \beta} \, \mathbf{g}_\alpha \mathbf{g}_\beta  \,,
\end{equation}
where $\mathbf{g}^\alpha$ are the contravariant base vectors, and covariant and contravariant components of vector $\mathbf{v}$ and tensor $\mathbf{T}$ are related as
\begin{equation}\label{eqA5}
v^\alpha = g^{\alpha \mu} v_\mu \,, \,\, T^{\alpha \beta} = g^{\alpha \mu} g^{\beta \nu} \, T_{\mu \nu}  \,.
\end{equation}

Since base vectors $\mathbf{g}_\mu$ are generally non-unit, vector components $v_\alpha$ or $v^\alpha$ are not coordinate invariant in curvilinear coordinates $x^\alpha$. Hence, they do not represent physical quantities. To obtain physically meaningful components of vectors, we have to substitute the base vectors $\mathbf{g}_\mu$ and $\mathbf{g}^\mu$ by normalized unit base vectors $\mathbf{e}_\mu$ and $\mathbf{e}^\mu$ \citep{Moller1972, Hartle2003}
\begin{equation}\label{eqA6}
\mathbf{e}_\mu = \mathbf{g}_\mu / \sqrt{g_{\mu \mu}} \,,\,\, 
\mathbf{e}^\mu = \mathbf{g}^\mu / \sqrt{g^{\mu \mu}} \,\,\, \mathrm{(no \, summation \, over} \, \mu) \,. 
\end{equation}
Consequently,
\begin{equation}\label{eqA7}
\mathbf{v} = v^{(\mu)} \mathbf{e}_\mu = v_{(\mu)} \mathbf{e}^\mu  \,. 
\end{equation}
where
\begin{equation}\label{eqA8}
v^{(\mu)} = v^\mu \, \sqrt{g_{\mu \mu}} \,, \,\, 
v_{(\mu)} = v_\mu \, \sqrt{g^{\mu \mu}} \,\,\, \mathrm{(no \, summation \, over} \, \mu) \,,  
\end{equation}
are the physical (proper) components of vector $v$. For orthogonal curvilinear coordinates, we get
\begin{equation}\label{eqA9}
v^{(\mu)} = v_{(\mu)} \,.  
\end{equation}

Another physically meaningful (proper) quantity is the infinitesimal distance in the Riemannian manifold defined as \citet[their eq. 13.3]{Misner1973}
\begin{equation}\label{eqA10}
ds^2 = g_{\mu \nu}\, dx^\mu dx^\nu \,,  
\end{equation}
being independent of the choice of the coordinate system $x^\alpha$. For static problems, when the Riemannian manifold is described by the orthogonal coordinates (time is independent of spatial coordinates), the proper distance in the Riemannian manifold reduces to the distance in the standard three-dimensional Euclidean space.


\end{document}